\documentclass[12pt,preprint]{aastex}

\newcommand{\be}{\begin{eqnarray}}
\newcommand{\ee}{\end{eqnarray}}

\newcommand {\nbody}{\textsc{\mbox{nbody6}}}
\newcommand {\nbodypp}{\textsc{\mbox{nbody6\raise.4ex\hbox{\tiny++}}}}
\newcommand {\COri} {\mbox{$\theta^1{\rm{C}}\:{\rm{Ori}}$}}
\newcommand {\rclose} {$r_{\rm{close}}$}

\newcommand {\Msun} {\mbox{M$_{\odot}$}}
\newcommand {\Lsun} {\mbox{L$_{\odot}$}}

\shorttitle{}
\shortauthors{Olczak, Pfalzner \& Spurzem}

\begin{document}

\title{Encounter-triggered disc mass loss in the ONC}
\author{C. Olczak, S. Pfalzner} 
\affil {I. Physikalisches Institut, University of Cologne, Germany} 
\author{R. Spurzem}
\affil {Astronomisches Rechen-Institut, Zentrum f\"ur Astronomie,
Univ. Heidelberg}

\begin{abstract}
The relevance of encounters on the destruction
of protoplanetary discs in the Orion Nebula Cluster (ONC) is investigated
by combining two different types of numerical simulation. First, star-cluster 
simulations are performed to model the stellar dynamics  of the ONC,
the results of which are used to investigate the frequency of encounters,  the 
mass ratio and separation of the stars involved, and the eccentricity of 
the encounter orbits. The results show that interactions that could 
influence the star-surrounding disc are more frequent  
than previously assumed in the core of the ONC, 
the so-called Trapezium cluster. Second, a parameter study of star-disc encounters 
is performed to determine the upper limits of the mass loss of 
the discs in encounters.  For simulation times of $\sim$\,1-2\,Myr 
(the likely age of the ONC) the results show that
gravitational interaction might account for a significant disc mass loss in dense 
clusters.  Disc destruction is dominated by encounters with high-mass stars,
especially in the Trapezium cluster, where the fraction of 
discs destroyed due to stellar encounters can reach 10-15\%.
These estimates are in accord with observations 
of \cite{lada:aj00} who determined a stellar disc fraction of 
80-85\%.
Thus, it is shown that in the ONC - a typical
star-forming region - stellar encounters do have a significant effect on 
the mass of protoplanetary discs and thus affect the formation of planetary systems.
\end{abstract}


\section{Introduction}
According to current knowledge planetary systems form from the accretion discs around young stars. 
These young stars are in most cases not isolated but are part of a cluster. It is still an open question 
as to how far encounters with the surrounding stars of the cluster influence planet formation. 
The fact is that these discs disperse over time either by photo-evaporation, encounter-induced 
disc mass loss or some other means. Simple calculations seem to indicate that encounters do not 
play an important role. Nevertheless, \cite{scally:mnras01} found that a little less than a third of stars in the central core of the ONC would suffer an encounter within 100\,AU, and thus, for a significant minority of stars in the Trapezium cluster, star disk encounters are of some importance. However, in order to quantify how many stars would be expected to lose most of their disc material, it is necessary to treat the disc mass loss - and its dependence on the mass of the perturbing object - in a more sophisticated way than was done by \cite{scally:mnras01}.

Like \cite{scally:mnras01} the ONC is used as model cluster for the following reasons: The ONC is 
thought to be a typical environment for star-formation and its high density suggests that stellar 
encounters might be relevant for the evolution of circumstellar discs. In addition, it is one of 
the best-studied regions in our galaxy, so that observational constraints significantly reduce 
the modelling parameters.
\subsection{Structure and Dynamics of the ONC}

The ONC is a rich stellar cluster with about 4000 members with masses $M^*$ above 0.08\,\Msun\ 
in a volume $\sim$ 5\,pc across \citep{hillenbrand:apj98,hillenbrand:apj00}. Most of the objects are 
T~Tauri stars, but there is also strong evidence for the existence of several protostars. The mean 
stellar mass is about 0.5\,\Msun\ \citep{scally:mnras05}, the half-mass radius roughly 1\,pc. 
Recent studies on the stellar mass distribution 
\citep{hillenbrand:apj00,luhman:apj00,muench:apj02,slesnick:apj04} reveal no significant 
deviation from the field star IMF \cite{kroupa:mnras93}
\be
\xi(M^*)= \left\{
\begin{array}{lrll}
0.035 M^*\,^{-1.3} & \mbox{ if } & 0.08&\le M^*\,<0.5, \\
0.019 M^*\,^{-2.2} & \mbox{ if } & 0.5&\le M^*\,<1.0, \\
0.019 M^*\,^{-2.7} & \mbox{ if } & 1.0&\le M^*\,<\infty. 
\end{array} \right.
\label{eq:imf}
\ee

The shape of the system is not perfectly spherical but elongated in the north-south direction. The reason for this asymmetry is the 
gravitational potential of a massive molecular ridge in the background of the cluster, OMC~1, being part of the much larger complex 
of the Orion Molecular Cloud. The mean age of the whole cluster has been estimated to be about 1-2\,Myr, though
with a significant age spread of the individual stars.
Today, star formation is no longer occurring in the cluster itself, only in the background molecular cloud.  
 After a short period of intense star-formation the ONC has expelled most of the residual gas by now.

The density and velocity distribution
of the ONC resembles an isothermal sphere. From the outer edge the number of stars falls linearly with decreasing radius $r$ down to 
$\sim$\,0.1\,pc; inside this cluster core, the distribution function becomes flatter 
\citep{jones:aj88,mccaughrean:aj94,hillenbrand:aj97,hillenbrand:apj98,mccaughrean:msngr02}.
The central number density $\rho_{\rm{core}}$ in the inner 0.053\,pc reaches $4.7\times10^4\,\mbox{pc}^{-3}$ \citep{mccaughrean:aj94} 
and makes the ONC one of the densest star forming regions in the Galaxy. The
velocity dispersion
is
nearly constant for all cluster radii. In their proper motion study of the ONC, \cite{jones:aj88} have obtained a three-dimensional velocity dispersion of $4.3\,$km\,s$^{-1}$,
 thus the crossing time is
$ t_{\rm{cross}}=2R_{\rm{hm}}/{\sigma}\approx0.5~\,\mbox{Myr}$.
%
%
Another crucial quantity in stellar dynamics is the virial ratio $Q_{\rm{vir}}$, 
which  has been estimated for the ONC as 
\begin{eqnarray}
Q_{\rm{vir}}=\frac{R_{\rm{hm}}\sigma^2}{2GM}\approx1.5~,
\end{eqnarray}
where $R_{\rm{hm}}$ is its half-mass radius. This indicates that the ONC is not 
only far from virial equilibrium, but even seems to be gravitationally unbound ($Q_{\rm{vir}}>1)$. 
However, this statement has to be treated with care because errors in the observational parameters 
can easily account for an error of over 50 per cent in this calculation. Besides, the estimated 
total mass of the ONC of 2000\,\Msun\ is only a lower limit since a substantial amount could be 
present in undetected low-mass binary companions or gas. Furthermore, the contribution of the 
OMC~1 to the overall gravitational potential is still unknown and the elongated shape of the 
cluster indicates that it is not negligible. However, as long as measurements of velocities and 
masses lack higher precision, one cannot constrain the cluster dynamics to contraction, equilibrium 
or expansion.

Like many other stellar aggregations, the ONC shows
 mass segregation, with the most massive stars being confined to the inner cluster parts. The Trapezium cluster, a subgroup of about 1000 stars in a volume 0.6\,pc across, represents the denser core of the Orion Nebula Cluster. It contains four luminous O and B stars at the very center, designated as the Trapezium.
Their most prominent member, \COri, is a O6 star with a mass of about 50\,\Msun, a luminosity of $4\times10^5\,\Lsun$ and a surface temperature of $4\times10^4\mbox{K}$.

Apart from its high density and young age, the evidence for protoplanetary discs around many 
stars in this cluster makes the ONC the ideal candidate for the present investigation. 
Whereas the first identification of ``peculiar stellar objects'' dates 
back to the paper from \citet{laques:aap79}, it took more than a decade to recognize 
them as circumstellar discs which are ionized by the intense radiation of the Trapezium stars.
\citet{o'dell:apj93} designated these bright objects as ``proplyds''. At greater distances from the cluster centre they also detected their dark counterparts: discs in silhouette which are visible due to the bright nebular background. 
Thus far, about 200 bright proplyds and 15 silhouette discs have been revealed in 
several HST studies of the ONC 
\citep{o'dell:apj93,o'dell:aj96,bally:aj98a,bally:aj00}, 
nearly all located in the Trapezium cluster due to selection effects. 

The most recent study on circumstellar discs in the Trapezium \citep{lada:aj00} used the 
L-band excess as detection criterium. They analyze 391 stars and find a fraction of 80-85\% to be 
surrounded by discs. This is in agreement with an earlier investigation of the larger ONC in which 
\citet{hillenbrand:aj97} states a disc fraction of 50-90\%. The disc sizes established so far 
vary between 50\,AU and 1000\,AU, with a typical value of 200\,AU for low-mass
stars. The inferred disc masses are only accurate to an order of magnitude but seem not to exceed a few percent of the central stellar mass,
which classifies them as low-mass discs. The disc surface densities are generally well described by power-law profiles, $\Sigma(r) \propto
r^{-a}$, with $0.5\le{a}\le1.5$.

\subsection{Previous work}

At present it is not possible to perform numerical particle simulations where the stars including
their surrounding discs are sufficiently resolved to determine the effect of encounters 
on the discs quantitatively. Therefore \cite{scally:mnras01} treated the dynamics of the stars in the
cluster and star-disc encounters and photoevaporation effects in separate investigations and combined
the results to determine the disc destruction rate. The term star-disc encounters
means here encounters where only one of the stars is surrounded by a disc, in contrast to disc-disc
encounters denominating encounters where both stars are surrounded by discs. 

In order to determine the disc destruction rate a number of assumptions were made by \cite{scally:mnras01}:

\begin{enumerate}
\item 
   The discs around the stars do not alter the stellar dynamics in any significant way.
\item 
   The discs are of low mass.
\item 
   Only two stars are involved in an encounter event and three-(or even more) body events are so rare that they can be neglected. 
\item The encounters were modelled as coplanar and prograde.
\item The disc mass loss is deduced from parameter studies of star-disc encounters.
\item 
The closest encounter is the most destructive one.
\item 
A parameter study of encounters between equal mass stars is used.

\end{enumerate}

From the last point
\cite{scally:mnras01} concluded that only in penetrating encounters a significant amount of mass can be stripped 
from the disc. Assuming a typical disc size of 200\,AU, they considered the stars with separations of
less than 100\,AU.
Under these assumptions \cite{scally:mnras01} found that just 3-4\% of all discs have the potential to be destroyed 
by encounters.

\subsection{Aim and structure of present work}

In this work we concentrate on the effect of encounters on the disc dispersal and ignore 
photo-evaporation. We performed an investigation similar to that of \cite{scally:mnras01} 
but drop the last two assumptions of above list.  In contrast to \cite{scally:mnras01} 
i) we record the entire path history of each star, so the effect of repeated encounters is included, 
ii) extend the parameter study to include encounters with massive perturbers and 
iii) record the most forceful encounter. It will be demonstrated that these modifications alter 
significantly the result, so that the conclusion that encounters can be 
excluded as a disc destruction mechanism should be revised. 

The paper is organized the following way: In Section~2 we begin with simulations of the ONC. This is  followed by an investigation 
of the mass loss in star-disc encounters in Section~3, resulting in a fitting formula for the disc mass loss. 
In Section~4 the results of Section~2 and 3 are combined to determine the disc dispersal as a function of time. 
In Section~5 it will be discussed how the disc mass loss is influenced by the assumptions made. 
\section{Cluster Simulations}

\subsection{Initial conditions}

The dynamical models of the ONC presented here contain only pure stellar components without considering 
gas or the potential of the background molecular cloud OMC~1.
All cluster models were set up with a spherical density distribution $\rho(r) \propto r^{-2}$ and a Maxwell-Boltzmann isotropic
velocity distribution. The masses were generated randomly according to the mass function 
given by Eq.~(\ref{eq:imf}) in a range $50\,\Msun\ge M^* \ge 0.08\,\Msun$ 
apart from the most massive star, representing \COri, which was directly assigned a mass of 50\,\Msun,
as in a random mass generation process only in very few cases the highest mass exceeds 
30\,\Msun.  
\COri\ was placed at the cluster centre and the other three Trapezium members were 
assigned random positions in a sphere of $0.3R_{\rm{hm}}$.
This procedure accounts for the initial mass segregation which is 
observed in young clusters and follows the study of \citet{bonnel:mnras98}. 
Except for the positioning of these three stars and the separate generation of \COri, this 
configuration is identical to the setup of \citet{scally:mnras01}.

The ONC was simulated for a total time of 13\,Myr, which is the assumed lifetime of \COri.
The quality of the dynamical models was determined by comparison to the observational data after a simulation time of 1-2\,Myr, which marks the range for the mean age of the ONC. The quantities of interest were the number of stars, the half-mass radius, the number densities, the velocity dispersion and the projected density profile.

The virial ratio $Q_{\rm{vir}}$ of the cluster is a crucial quantity for its dynamics.
Contracting models ($0.01\le Q_{\rm{vir}}<0.5$) showed only moderate agreement 
with observations in the density distribution and only after at least 2\,Myr simulation time. However, this scenario was not excluded for the study since the age spread of the ONC is large and star formation set in more than 4\,Myr ago.
Unbound models with $Q_{\rm{vir}}>1.0$ were not further considered since unreasonable high initial core densities of $\rho_{\rm{core}}>10^6\,{\rm{ pc}}^{-3}$ are required in this case and the resulting density distribution did not resemble the observations very well. Hence, only models with $0.01\le Q_{\rm{vir}}\le 1.0$ are presented here. This constraint is also in accordance with the recent results from \citet{scally:mnras05}.


Three configurations, A, B and C, were chosen as the best dynamical models of the ONC, according to $Q_{\rm{vir}}=0.5$, $Q_{\rm{vir}}=1.0$ and $Q_{\rm{vir}}=0.1$, respectively. 
The parameters of the three configurations are summarized in Table~\ref{table:sim2}.
Ten random setups of each of these initial models were simulated to minimize statistical variations in the results.

%
\subsection{Numerical method}
The cluster simulations were performed with \nbodypp\ \citep{spurzem:mnras02}, which allows 
a high-accuracy treatment of two-body interactions and is a parallelized version of \nbody\ 
\citep{aarseth:book03}. Since the aim of this work was to record the encounters in the ONC, 
additional routines had to be implemented:

Apart from storing the closest encounter for each star a more advanced encounter list was produced: 
The search criterion for the next perturber of a star was modified by considering the gravitationally 
most dominating body instead of the closest neighbour, or in other words, the minimization of the distance 
was replaced by the maximization of the gravitational force.
%
The former scheme underestimates the effect of stellar encounters since the nearest neighbour is not 
necessarily the gravitationally dominant one.
Approaches of stars were only considered to be true encounters if the orbit of the perturber was concave and if
the calculated relative disc mass loss was higher than the $1\sigma$ error, which is 0.03 in this study. 

As the above approach would only account for one single encounter for every star, the encounter 
list was extended by recording the information of {\em all} perturbing events of each stellar 
disc during the course of the simulation.
In order to obtain the disc mass loss,  both masses, the relative velocity and the eccentricity 
were additionally recorded, which constitute the full set of orbital parameters determining the 
planar two-body problem.

%
%
%

\subsection{Cluster results}

\begin{figure}
\epsscale{1.0}
\plotone{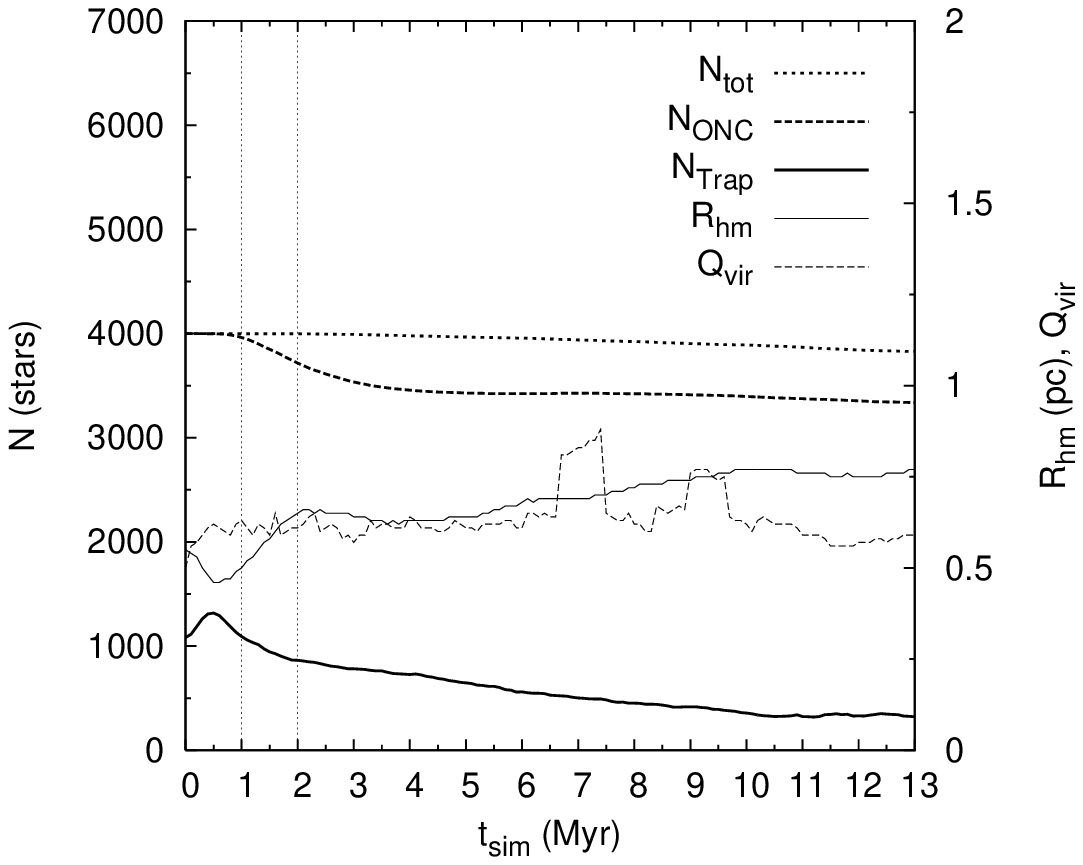}
\caption{Time evolution of particle numbers, half-mass radius and virial ratio
  of the model cluster A. The particle numbers represent the total population,
  $N_{\rm{tot}}$, the population of the ONC ($R_{\rm{ONC}}=2.5\,$pc),
  $N_{\rm{ONC}}$,  and that of the Trapezium cluster
  ($R_{\rm{Trap}}=0.3\,$pc), $N_{\rm{Trap}}$, respectively.
\label{evol}}
\end{figure}

Here we will mainly describe the results from the model in virial equilibrium (A) and discuss the
expanding and contracting model in Section 4. In Fig.~\ref{evol} the temporal development of the total 
particle number and the number of particles confined to the volume of the ONC and the Trapezium, 
are shown. The total particle number in the simulation decreases because
particles reaching the numerical cutoff radius ($\sim 20 R_{\rm{hm}}$) are excluded from the simulation.
The particle number in the entire simulation volume is nearly unchanged for the whole simulation time 
of 13\,Myr, only 4\% of all particles exceeded the cutoff radius. If one considers the 
ONC volume, the population is reduced slightly more, but only for simulation times larger than 2\,Myr, 
marked by the right vertical dotted line. Still 95\% of the initial number is preserved and thus satisfies the 
observational constraints. In the Trapezium volume, the population initially rises, 
which means that the inner core of the cluster undergoes some contraction. After
approximately 0.5\,Myr this trend is inverted, reducing the stellar number to nearly 1000 or 800 
after 1\,Myr or 2\,Myr, respectively. This behaviour is also
reflected in the time-dependence of the
half-mass radius on the right-hand axis, which decreases from 
0.55\,pc to 0.46\,pc at its minimum.

The rise of the virial ratio at the beginning of the simulation is mainly due to the gain of potential 
energy by contraction of the cluster, which results in a rise of the kinetic energy by an equal amount 
due to energy conservation. The heating of the entire core finally overcomes the gravitational drag and 
inverts the inward motion of the stars; the fastest of these soon leaving the boundary of the ONC. After 
several crossing times the system begins to relax and  the virial ratio is roughly constant, 
$Q_{\rm{vir}}=0.6$, which means that the cluster is still a bound entity. 

\begin{figure}
\epsscale{1.0}
\plotone{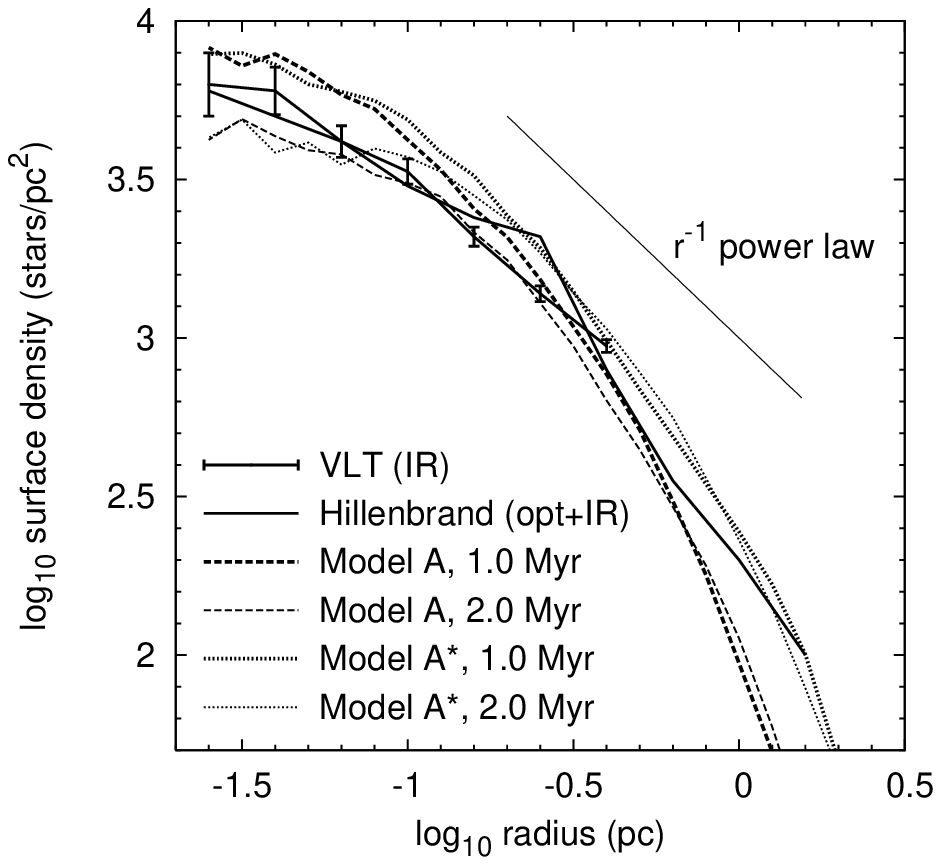}
\caption{The density profile of the model cluster A at $t_{\rm{sim}}=1$\,Myr
  and $t_{\rm{sim}}=2$\,Myr is compared to data from infrared observations
  performed by \citet{mccaughrean:msngr02} with the VLT. The original figure
  is from \citet{scally:mnras05} and includes two additional data sets from
  \citet{jones:aj88} and \citet{hillenbrand:aj97}. The set according to
  \citet{jones:aj88} has been excluded from this image for greater
  clearness. For comparison a power law of slope -1 is shown, corresponding to
  $r^{-2}$ in three dimensions, the estimated profile of the ONC. The
  equivalent distributions for the simulations by \cite{scally:mnras01} are
  denoted as model~A$^*$. Here, due to the lower density of model~A$^*$ by a factor of two, the distributions were shifted by the same factor to allow for comparison.
\label{dens:dist}}
\end{figure}

In Fig.~\ref{dens:dist} the density profile of the model cluster A at 1\,Myr and 2\,Myr is compared to 
data from infrared observations performed by \citet{mccaughrean:msngr02}. Both curves fit well up to a 
distance of 0.3\,pc, and therefore qualify as a model for the ONC. For larger distances the values are 
somewhat below the observed profile. This is due to the initially truncated particle 
distribution and evaporation of the system. However, since this investigation focuses on the effect of encounters 
on the disc mass in the inner Trapezium cluster, this should not significantly influence the results.
%
%

\begin{figure}
\epsscale{0.6}
\plotone{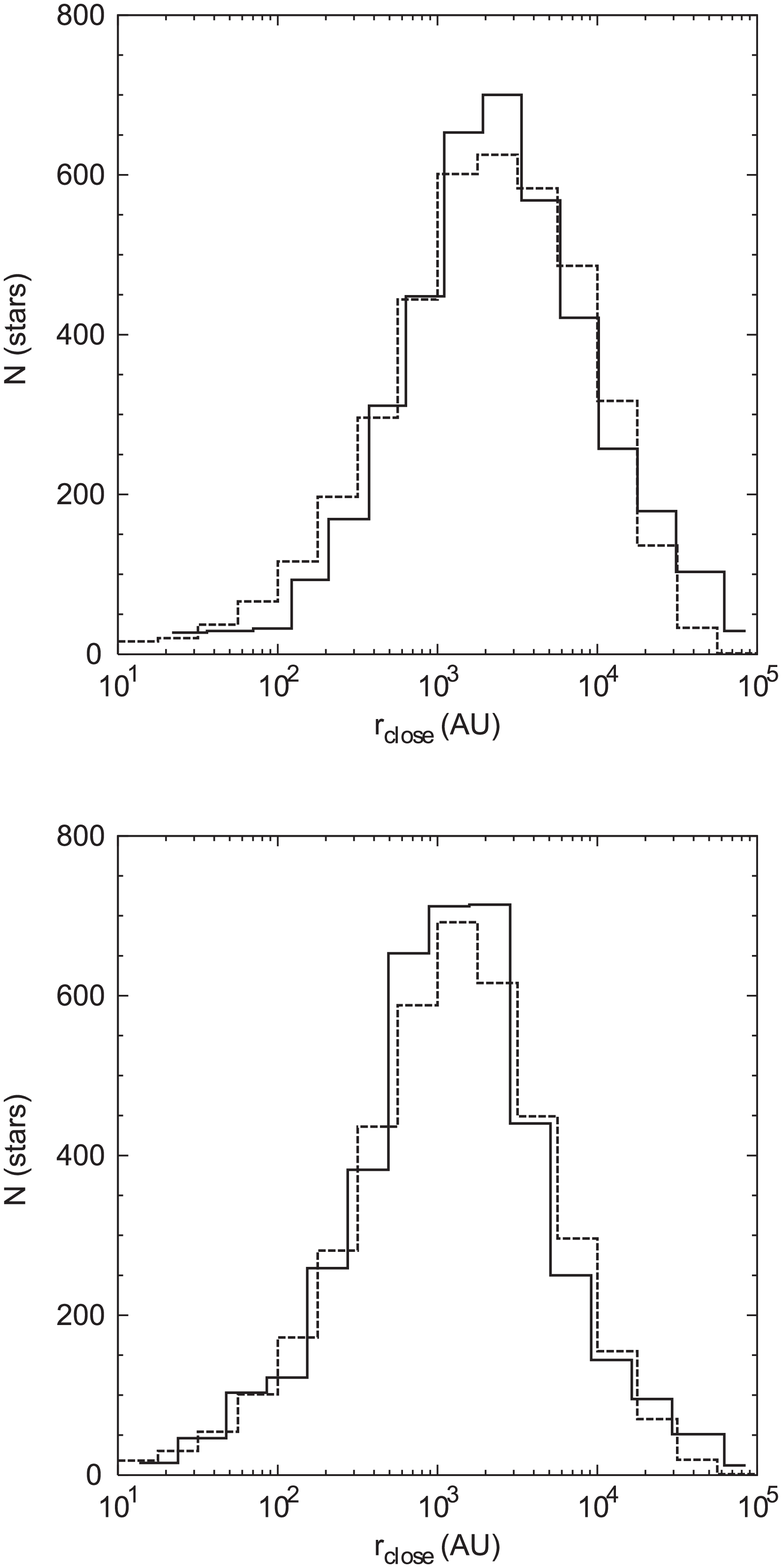}
\caption{Number of stars as a function of the closest encounter
  distance. Comparison of model~A$^*$ (dashed lines) with that of
  \cite{scally:mnras01} (drawn lines) at a) $t_{\rm{sim}}=2.9$\,Myr and b)
  $t_{\rm{sim}}=12.5$\,Myr.
\label{hist:dist}}
\end{figure}

As mentioned before one main reference was the work of \citet{scally:mnras01} who found stellar 
encounters to play only a minor role even in such a dense environment. The initial conditions of their 
model (designated as model~A*) are identical to model~A with the exception of a nearly two times larger 
initial extension. As Fig.~\ref{hist:dist} shows, choosing an analogous 
cluster setup, the distribution of closest encounters presented in \citet{scally:mnras01} could be 
reproduced very well.
Fig.~\ref{hist:dist} presents two histograms of \rclose\ for all the stars in the cluster after a 
simulation time of 2.9\,Myr and 12.5\,Myr, respectively, with \rclose\ for each star representing 
its overall minimum separation to any other object in the cluster - this quantity can only either 
remain constant or decrease with time. At a simulation time of 2.9\,Myr only few ($\sim4\%$) very 
close encounters with a separation of less than 100\,AU have occurred, while the majority of stars  
never approached closer than 1000\,AU 
and even after 12.5 Myr  only slightly more than 5\% of the stars were encountered closer than 100\,AU.



%
%
%
%
While the shape of the projected density profile shown in
Fig.~\ref{dens:dist} provides a reasonable fit to the observational data, the magnitude of the density in the inner cluster parts
is nearly two times lower (thus the density profile had to be shifted upwards in figure~\ref{dens:dist} for comparison with the VLT data). 
So that we conclude that model~A provides a better fit to the observational data and will
apply it in the analysis that follows.

\begin{figure}
\epsscale{0.6}
\plotone{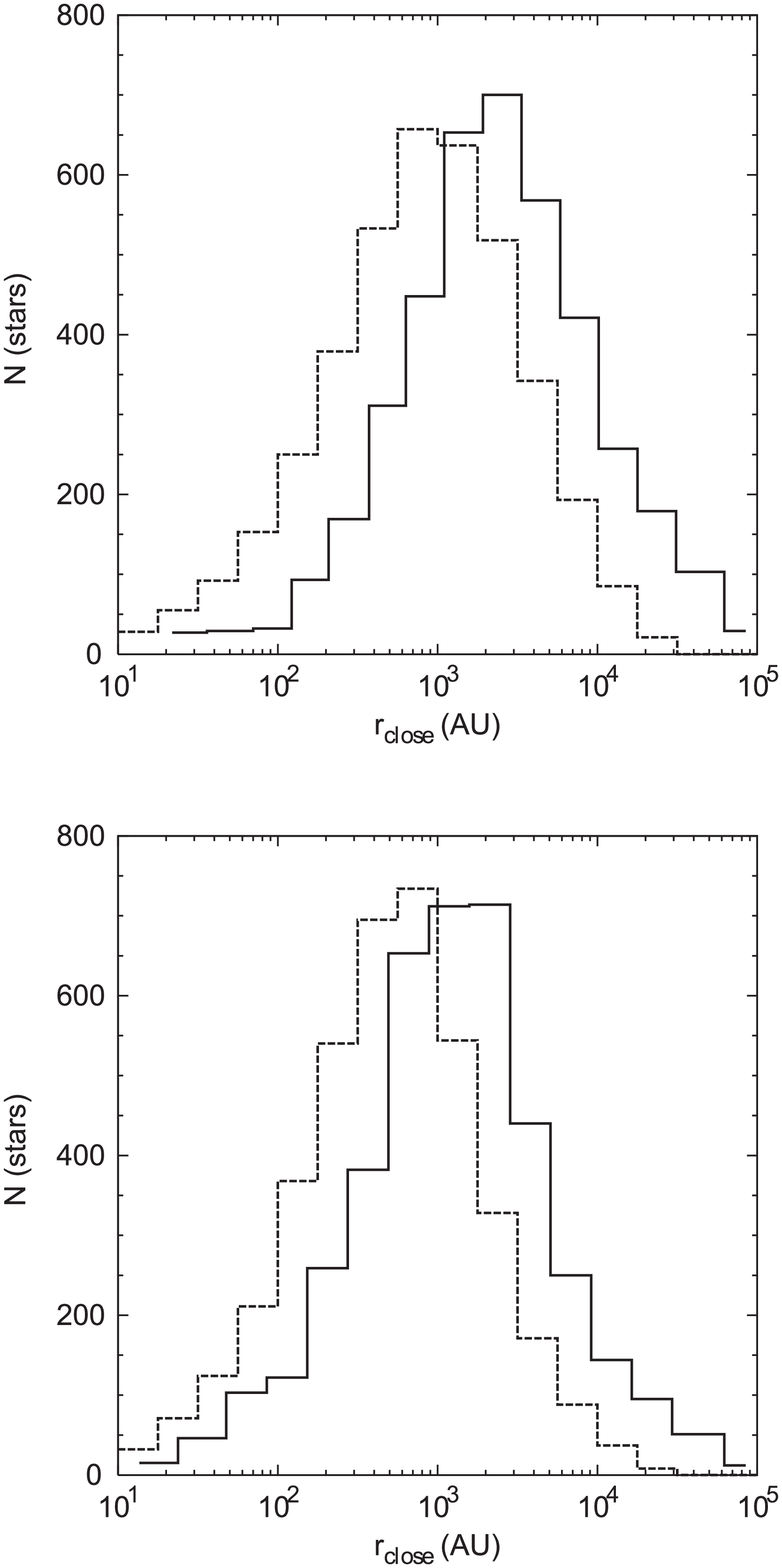}
\caption{Number of stars as a function of the closest encounter
  distance. Comparison of model~A (dashed lines) with that of
  \cite{scally:mnras01} (drawn lines) at a)  $t_{\rm{sim}}=2.9$\,Myr and b)
  $t_{\rm{sim}}=12.5$\,Myr.
\label{hist}}
\end{figure}

Fig.~\ref{hist} demonstrates that for model~A the median is far below 1000\,AU, after 12.5\,Myr 
it even approaches 500\,AU. This shifting 
directly reflects the roughly two times higher initial density of model~A. Analogously, the number 
of stars with \rclose\ lower than 100\,AU more than doubles resulting in a fraction of 9.4\% for 
2.9\,Myr and 11.9\% after 12.5\,Myr.

%

\subsection{Mass Loss}

Combining the results of Section 2 and 3, the relative disc mass loss for each disc is obtained 
as a function of the simulation time. The parameters of each stellar encounter were taken from 
the stellar encounter lists  generated by preprocessing the original encounter list 
of each single run. The calculation of the disc mass loss due to encounters was performed for 
each single run and then averaged over ten simulations.

According to the improved fit function~(\ref{eqn:ImprovedFit}), the relative
mass loss of a stellar disc was obtained from the perturber periastron. For this purpose, first
the disc size of the central star was scaled due to the stellar mass by using Eq.~(\ref{eqn:rdisk}), 
then the relative perturber mass, $M_2^*/M_1^*$, and the relative periastron, $r_{\rm{p}}/r_{\rm{d}}$ were passed to 
the fit function~(\ref{eqn:ImprovedFit}). The errors of the estimated
relative disc mass loss due to each encounter $i$, $\Delta^i=\Delta^i(r_{\rm{p}}/r_{\rm{d}})$, were 
assumed to be $\Delta^i=0.03$ for $r_{\rm{p}}/r_{\rm{d}}>1$, $\Delta^i=0.05$ for $r_{\rm{p}}/r_{\rm{d}}>0.1$ and 
$\Delta^i=0.1$ if $r_{\rm{p}}/r_{\rm{d}}=0.1$, according to the statistical errors of the encounter simulations.

In the following it will be shown for our model~A that an improved encounter treatment is important 
because (a) the majority of the stars in the model clusters undergo more than one encounter, (b) a large 
fraction of the stars encounter a much more massive perturber, and (c) the largest perturbation of a disc 
is caused by the gravitationally most dominating body and \textbf{not} by the closest companion.

%
  
\begin{figure}
\epsscale{0.7}
\plotone{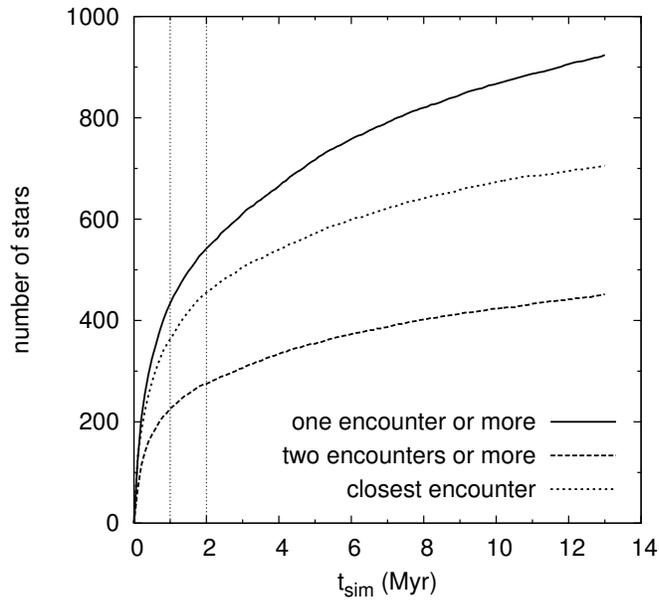}
\caption{Number of stars that were subject to an encounter as a function of
  the simulation time. Comparison of three different encounter scenarios: all
  stars with at least one encounter, only stars with repeated encounters, and
  stars with the closest encounter only.
\label{enc:num}}
\end{figure}

%
%
Defining an encounter as a perturbing event in which a circumstellar disc looses at least 3\% of its mass,
Fig.~\ref{enc:num} shows that the effect of (c) for the number of perturbed discs is only minor. In contrast, the 
improvement (a) has a much larger effect: roughly half of all discs that have been 
perturbed have been done so repeatedly.
Thus, a realistic investigation of stellar encounters in young clusters and their effect on 
protoplanetary discs cannot be performed without considering the entire encounter history of each star.

\begin{figure}
\epsscale{1.0}
\plotone{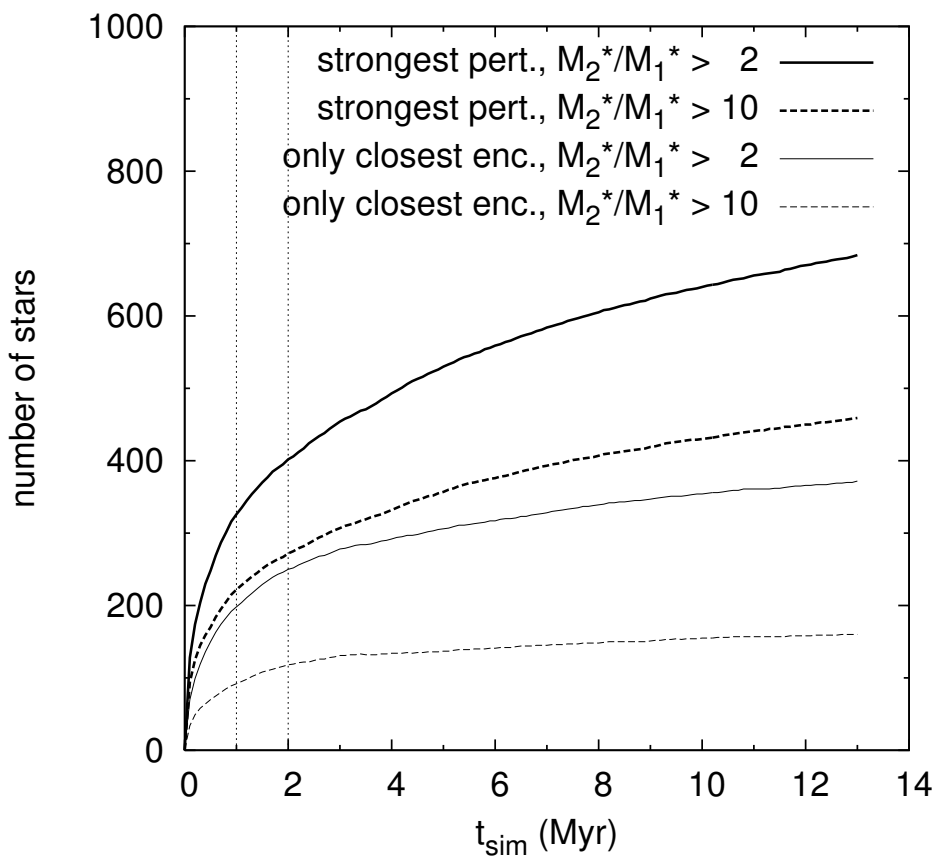}
\caption{Number of stars that were subject to an encounter as a function of
  the simulation time. Comparison of the two different encounter definitions
  used in this work. The effect on the relative perturber mass is shown
  exemplary for $M_2^*/M_1^*>2$ and $M_2^*/M_1^*>10$, respectively.
\label{enc:mass}}
\end{figure}

%
%
In Fig.~\ref{enc:mass} the effect of (b) and (c) is shown by plotting the ratio of the perturber mass $M_2^*$ and that of the encountered star $M_1^*$, $M_2^*/M_1^*=2$ and $M_2^*/M_1^*=10$, for the two encounter criteria, respectively. If this quantity is significantly higher than unity, then even non-penetrating stellar encounters have the potential to remove a large fraction of a disc's mass. In contrast to Fig.~\ref{enc:num} it is apparent that (c) has a major effect: this criterion causes a twice as large number of high mass ratios in both cases, $M_2^*/M_1^*=2$ and $M_2^*/M_1^*=10$, when comparing the strongest and the closest perturber. This means that while the number of perturbed discs does not change significantly, the effect of (c) on the disc mass loss is large. The effect of (b) is given by comparing the numbers with Fig.~\ref{enc:num}. It is evident that most perturbed discs have been approached by a companion at least twice as massive as the central star, and nearly half of them was more than ten times heavier. Again, this improvement of the treatment of encounters has a strong impact on the disc mass loss. 

\begin{figure}
\epsscale{0.4}
\plotone{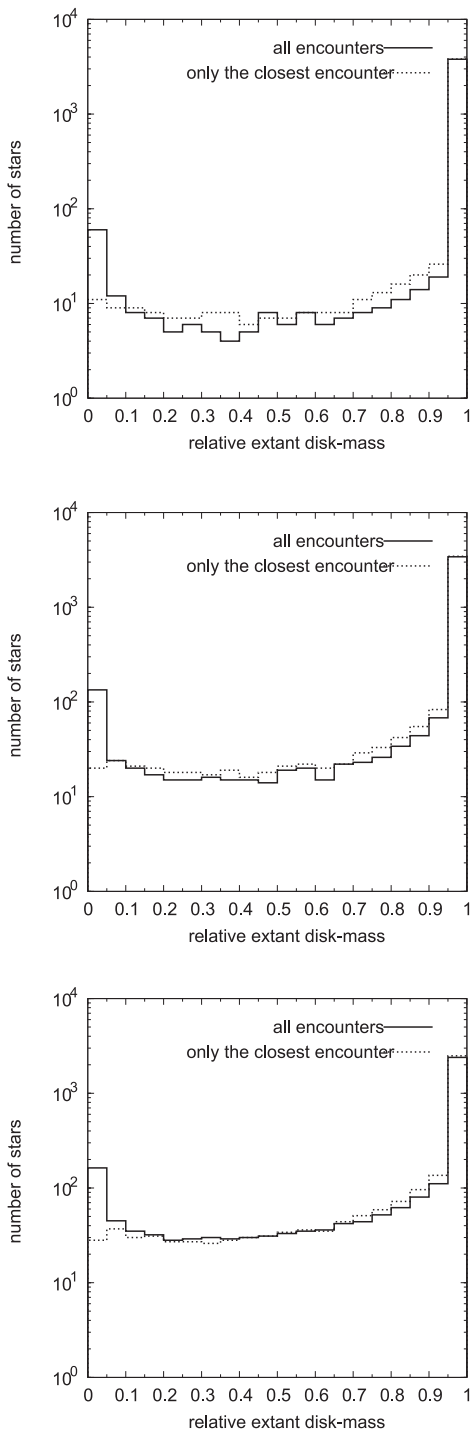}
\caption{Distribution of the relative extant disc mass of all discs at three
  different times for model~A. The output times correspond to shortly after
  the start of the simulation, $t_{\rm{sim}}=0.1\,$Myr, to the lower limit on
  the mean age of the ONC, $t_{\rm{sim}}=1.0\,$Myr, and to the end of the
  simulation at $t_{\rm{sim}}=13.0\,$Myr.
\label{lossdist1}}
\end{figure}

%
For the clusters in virial equilibrium (Model~A), Fig.~\ref{lossdist1} depicts the 
distribution of the relative disc mass at different simulation times in a histogram exemplary 
for an intermediate standard disc size, $r_{\rm{sd}}=r_{\rm{d}}(1\,\Msun)=150$\,AU. Each plot contains two 
distributions, one in which the entire encounter history of all stars in the cluster was considered, 
and the other accounting only for the closest encounter.
It is apparent from the plots that both distributions have a prominent minimum near relative 
disc masses of 0.5, which means that shortly after the simulation start the stellar population 
divides roughly into two groups, one containing the stars which have virtually undisturbed
discs and the other those which have lost the major part of their disc. The former diminishes from 
roughly  95\% at 0.1\,Myr and to 72\% at 
13\,Myr. There are in general only slight differences between the two distributions
of Fig. \ref{lossdist1} if one neglects the leftmost bin, which means that as long as the relative 
disc mass does not fall below 5\%, the closest encounter of each star
determines entirely its disc mass loss, independent of eventual additional encounters.

The situation is very different if one considers only the stars with at least a disc mass loss above 95\%. 
There are roughly six times more stars when one models repeated and high $M_2^*/M_1^*$ encounters,
more precisely 1.5\% instead of 0.25\% at 0.1\,Myr, and 4.9\% instead of 0.8\%
at 13\,Myr, respectively. So only a sequence of several encounters has the potential to
completely disrupt a significant number of protoplanetary discs. Hence, the encounter scenario used in 
this work decisively affects the statistics on the disc mass loss due to encounters, particularily 
for the extreme case of a nearly total removal of a star's disc material.

\begin{figure}
\epsscale{0.6}
\plotone{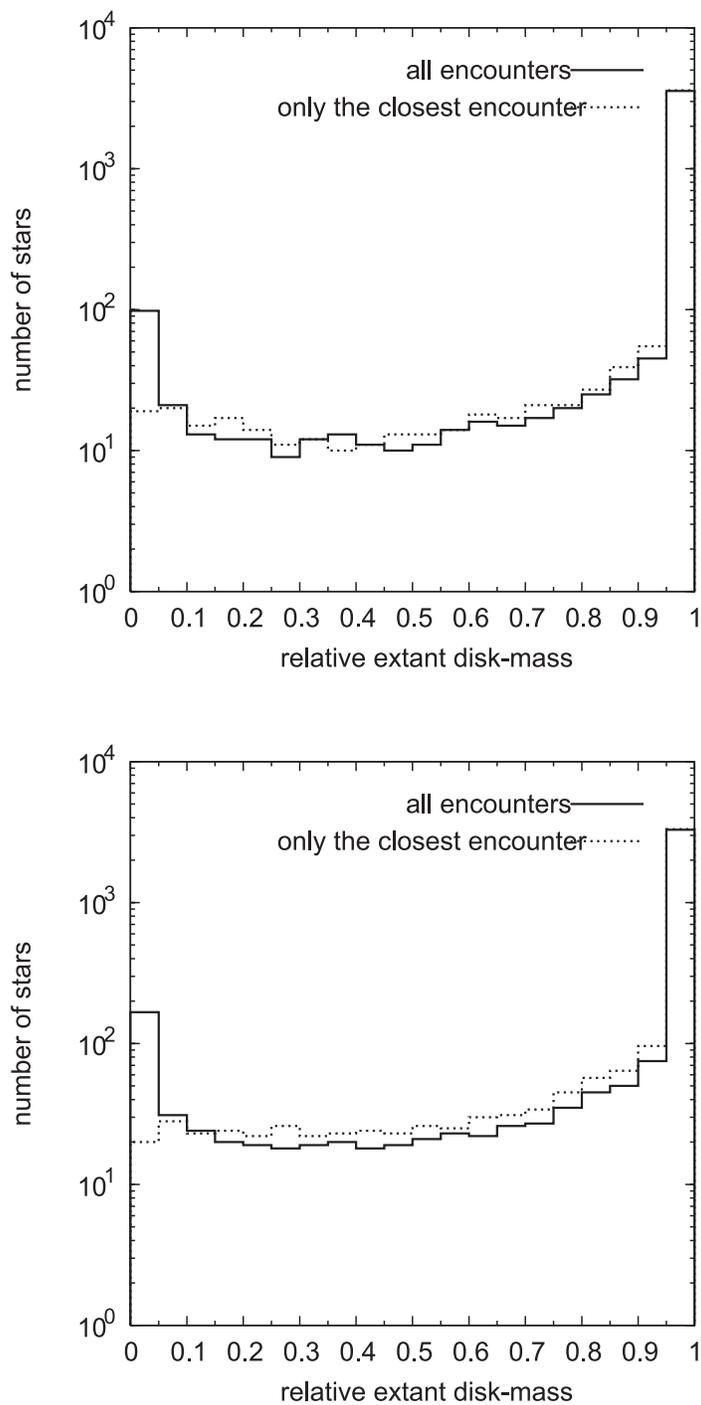}
\caption{Distribution of the relative extant disc mass of all discs after
  $t_{\rm{sim}}=1$\,Myr for two different standard disk sizes,
  $r_{\rm{sd}}=100\,$AU (top) and $r_{\rm{sd}}=200\,$AU (bottom), model~A.
\label{lossdist2}}
\end{figure}

\begin{figure}
\epsscale{0.6}
\plotone{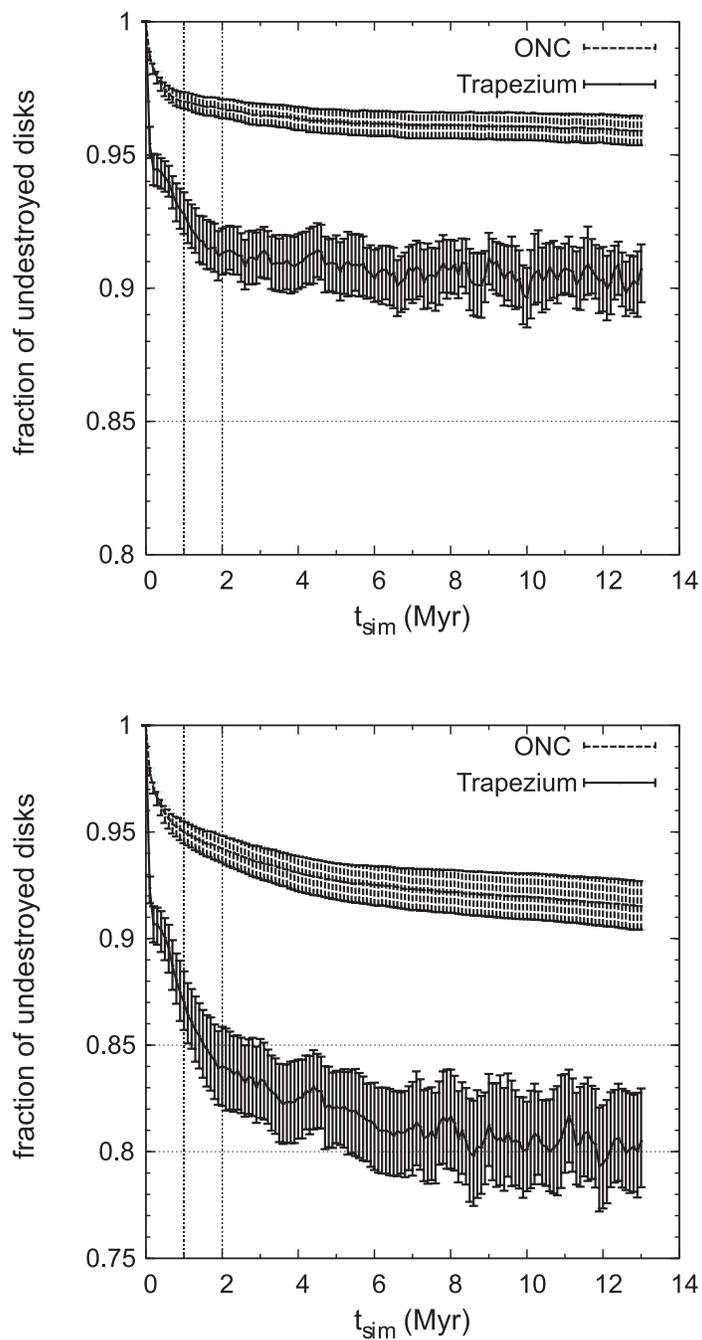}
\caption{Fraction of un-destroyed discs (i.e. discs with at least 10\% of
  their initial mass) in the ONC and the Trapezium cluster as a function of
  the simulation time, exemplary shown for $r_{\rm{sd}}=100\,$AU (top) and
  $r_{\rm{sd}}=200\,$AU (bottom), respectively, model~A.
\label{losstime}}
\end{figure}

In the second realization of the distribution of the relative disc mass, shown in 
Fig.~\ref{lossdist2}, the histograms are plotted for 
two different standard disc sizes, 100\,AU and 200\,AU, at
1\,Myr simulation time. It is apparent that in accordance with the doubled size of the standard disc size  
the disc mass loss is far higher. This finding demonstrates the sensitivity of the
disc mass loss on the choice of the standard disc size  $r_{\rm{sd}}$, the relation of which to the disc size
$r_{\rm{d}}$ is given by Eq.~(\ref{eqn:rdisk}),
and thus justifies 
its application as a free parameter as well in Fig.~\ref{losstime}.
%
%

Here, the fraction of still un-destroyed discs is plotted against simulation time for the same two standard disc sizes $r_{\rm{sd}}$.
The plot in Fig.~\ref{losstime} depicts the fraction of un-destroyed discs for the entire ONC and for the Trapezium cluster. The dotted horizontal line represents the upper limit of the fraction of disc surrounded stars in the Trapezium of 80-85\% \citep{lada:aj00}. It is apparent that the period of most violent disc destruction lasts for roughly 2\,Myr after the simulation start, consistent with the phase of slight contraction of 
model~A.

Equally, the doubled standard disc size affects the fraction of un-destroyed discs directly in the sense that the fraction of \emph{destroyed} discs becomes two times higher.

A comparison of the resulting fraction of remaining discs in the Trapezium with the observational 
estimates of \citet{lada:aj00} shows that 
the upper limit of 85\% is not surpassed 
in the specified range of the ONC's mean age. This is interpreted as a
correspondence with observations since it is likely that processes other than star-disc encounters, 
e.g. photo-evaporation \citep{scally:mnras01}, do account for the destruction of protoplantery discs 
and thus the fraction of remaining discs due to encounters
alone should be higher than the observational value. Moreover, the observational estimates state perhaps 
even lower limits since it is much more likely to fail with the detection of circumstellar discs due to 
sensitivity limits than to classify a bare star as a star-disc system. However,
with increasing time the fraction of remaining discs decreases further and reaches nearly an asymptotical 
value at the end of the simulation which implies that star-disc encounters in a dense core like the 
Trapezium could even destroy up to 20\% of all discs.

Considering now an expanding (model B) and contracting (model C) cluster, our simulations show
that for the expanding model in the Trapezium cluster the fraction of remaining discs rises
much faster due to the twice as high initial core density. However, the period of violent interaction 
is much shorter due to the fast expansion and so that again the fraction of un-destroyed discs 
is 85-95\% in the Trapezium but this level is already reached at 0.5 Myr.
However, in the case of the entire ONC, the fraction of remaining discs stays much higher than in 
model~A as here the fast expansion of the outer regions is not compensated by a highly increased density, which is required for close interactions of objects with high velocities.

In the contracting model the fraction of destroyed discs in the entire ONC is fairly low for all times.
For the Trapezium cluster the population grows faster than discs are destroyed so that 
the relative number of destroyed discs actually decreases. 
If the ONC was initially in a dynamically cold state, so that 90-95\% of the discs in the Trapezium 
cluster would not have been seriously disrupted and the effect of star-disc encounter on the disc mass 
loss of protoplanetary discs would be negligible in the case of a contracting cluster.

\section{Discussion}
\label{sec:discussion}

In the previous sections it was repeatedly stated that the results here represent an upper
limit for the destruction of discs by encounters in the ONC. In the following this will be
explained in more detail and estimates for lower limits of the mass loss given.

The situation described above contains, like previous work, a contradiction - it was assumed that
each star is initially surrounded by a disc and at the same time the disc mass
      loss that a star (without disc) produces in a star disc system was considered. 
      Logically both stars would have to be surrounded by a disc each. There are two reasons why
      this was done. First, encounters where both stars are surrounded by discs are
      less well investigated and second \cite{pfalzner:apj05} showed that the
      star-disc results can be generalized to disc-disc encounters as long as there
      is no mass exchange between the discs. In the case of a mass exchange the discs
      can be to some extent replenished so that the mass loss would be overestimated. 

\begin{figure}
\epsscale{1.0}
\plotone{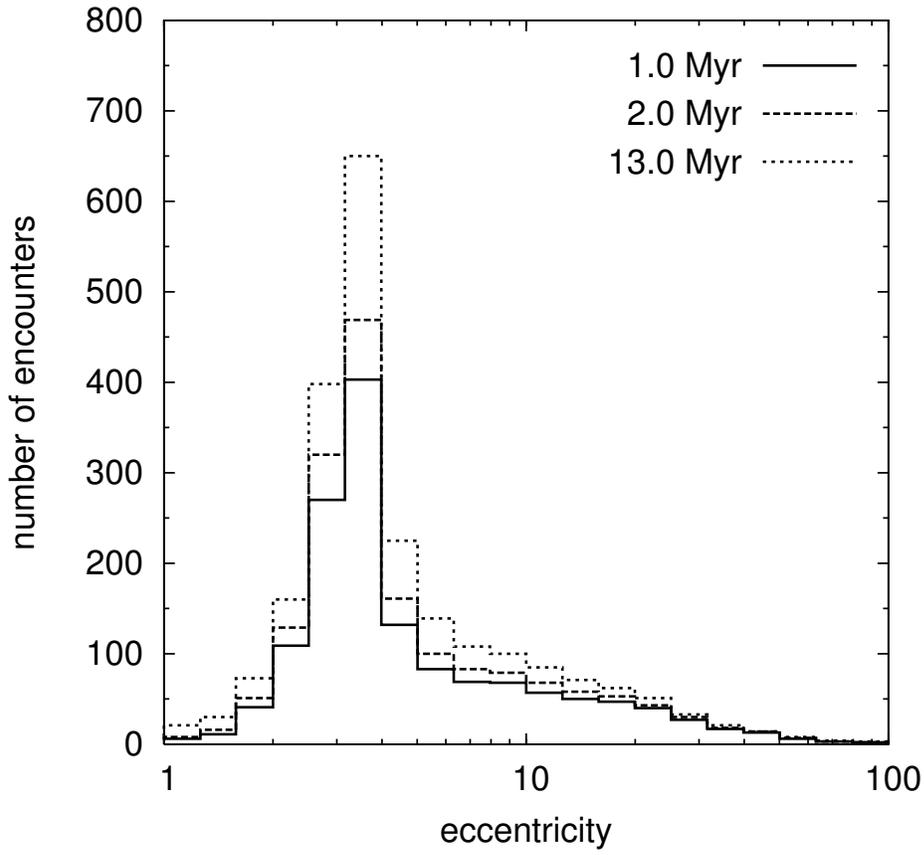}
\caption{Histogram of the eccentricity of all encounters which have led to the
  disruption of discs (i.e. removal of more than 90\% of the initial disc
  mass) within the specified simulation time, model~A.
\label{enc:ecc}}
\end{figure}

Eq.~(\ref{eqn:ImprovedFit}) is valid for parabolic encounters. Binary formation $(\epsilon < 1)$ can 
      be neglected as in this study it happens in less than 0.5\% of all encounters.
      However, most encounters in the cluster simulations are not parabolic but hyperbolic. In such 
      hyperbolic 
      encounters the mass loss is lower because the disturber is not long enough in the
      vicinity of the star-disc system to remove disc mass. However, considering only
      the stars that lose more than 90\% of their disc mass, the eccentricity $\epsilon$ of their orbits 
      has a maximum at $\epsilon \approx 3$ (see Fig.~\ref{enc:ecc}).
      \cite{pfalzner:aap05} showed that for $M_2^*=1\,\Msun$ the relative disc mass loss in an
      $\epsilon = 3$ encounter is about 55\% of that of a parabolic encounter.   
      
It is still an open question whether discs in clusters are in any way aligned and whether there
      is a preference for coplanar or prograde encounters due to the common formation history of the involved 
      stars and discs. If the coplanar, prograde encounters considered here are in any way favoured, they are
      the most destructive type of encounter.
However, in a cluster that is not highly flattened it seems rather unlikely that the encounter 
planes are to a high degree aligned. Therefore one would expect most encounters to be 
noncoplanar. \cite{pfalzner:aap05} showed that, as long as the inclination is not larger than  
45$^\circ$ the mass loss in the encounter is only slightly reduced in comparison to a coplanar 
encounter. If however the orientation is completely random and a 90$^\circ$-encounter the
most likely encounter scenario, the mass loss could be significantly reduced. This point needs 
further investigation.

In this investigation it was assumed that in repeated encounters the relative mass loss is the same.
      This is somewhat in contrast to the prevailing view that an encounter 'hardens' a disc so that
      later encounters can no longer influence it. This effect of disc hardening might happen for high-mass 
      discs but for low-mass discs \cite{pfalzner:apj04} found that although the total disc mass loss was smaller 
      in a second encounter the relative disc mass loss was the same as in the first encounter. 
      As this has only been tested in a particular case ($M_2^*= 1\,\Msun$), it should be investigated whether 
      this holds generally.
It is important to note that the calculations of \cite{pfalzner:apj04} are ballistic particle simulations, neglecting the effects of viscosity. Simulations of star-disc encounters by \cite{clarke:mnras93} have shown that in this case discs get puffed up by encounters. On the other hand, the inclusion of dissipation lead to recircularisation of the remnant disc and hence disc shrinkage. Thus one might conclude that disc hardening must be underestimated by \cite{pfalzner:apj04}. However, \cite{pfalzner:apj05} have as well investigated the effect of viscosity on different disc parameters and found no significant differences neither in the mass loss nor in the density distribution.

In this study it was assumed that all encounters can be described as two-body processes.
      \cite{umbreit:phd05} found from three-body encounters simulations that the resulting discs
     are flatter and less massive than after similar two-body encounters with the same minimum 
      encounter distance. 

In this work it has not been considered that a considerable proportion of the stars
      in the ONC are not single stars but binary systems. Further studies would be needed to
      see if binary systems would lead to a different disc destruction rate. A second point for
      future investigations would be the inclusion of gas in the ONC simulations.

%

One should as well emphasize that the disc mass loss falls below $10\%$ for 
mass ratios below 0.1, so the contribution from light perturbers is 
negligible as long as they are not penetrating the disc. 
Regarding the setup of the cluster simulations, the cut-off in the cluster 
IMF at substellar masses (M$^* \le 0.08\,\Msun$) should thus have only a 
minor effect on the cumulative disc mass loss.

\section{Conclusion}

This investigation for the ONC cluster, combining cluster simulations with encounter investigations, shows that potentially up to 10-15\% of the discs in the Trapezium cluster
could have been destroyed by encounters. Our more sophisticated treatment of disc mass loss expected from multiple stellar encounters implies that it is plausible that the 15-20\% of discless stars observed in the Trapezium \citep{lada:aj00} may, in a large fraction of cases, result from star-disc collisions.



One important result is that the most massive bodies dominate the disc mass loss, with significant interaction even beyond a separation of ten disc radii for a ONC-like entity. This is particulary so for the Trapezium, where some dozen massive stars are surrounded by hundreds of lighter bodies. Consequently, it is the upper end of a cluster's mass distribution that to a large degree determines the fate of the circumstellar discs in its vicinity and thus there are in principle two quantities that are mainly regulating the effect of stellar encounters on the mass-loss from protoplanetary discs: namely the local stellar density which determines the encounter probability, and the upper limit of the mass range, which affects the maximum strength of the perturbing force.

As the number of massive members in a stellar group seems to be correlated to its initial density \citep[see][]{testi:aap97,bonnel:mnras04} and the IMF appears to be uniform for all Galactic environments \citep[e.g][]{kroupa:mnras93,muench:apj00} of star-formation, the dependency of the disc mass loss due to encounters is mainly reduced to one parameter, namely the density distribution of the considered stellar system. This relation would be worth investigating in future.

As the possible disc destruction rate is so high, it is obvious that the remaining discs are considerably affected by encounters - that is the mass distribution and the cut-off radius of the disc. These properties are two ingredients vital for the understanding of the formation of planetary systems.  Further studies of the temporal evolution of these properties in the ONC are on the way.

\section*{Acknowledgments}
Simulations were partly performed at the John von 
Neumann Institut for Computing, Research Center J\"ulich, Project HKU14.
We want to thank the referee C.Clarke for her very useful comments.
RS wants to thank Sverre Aarseth for a sustained friendly and indispensable
collaboration over many a project and code.
\bibliographystyle{apj}

\clearpage

\clearpage

\begin{figure}
\epsscale{0.7}
\plotone{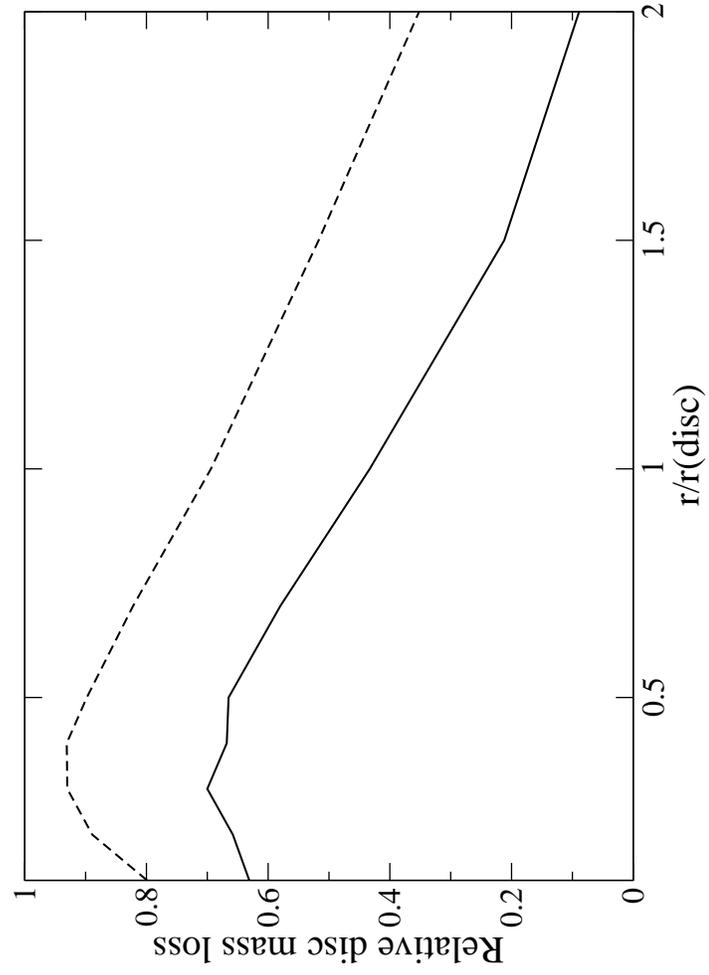}
\caption{Relative disc mass loss, $\Delta{M}_{\rm{d}}/M_{\rm{d}}$, in
  parabolic encounters as a function of the relative periastron distance,
  exemplary for $M^*_2=1\,\Msun$ (drawn line) and  $M^*_2=5\,\Msun$ (dashed
  line).
\label{fig:massloss}}
\end{figure}

\begin{figure}
\epsscale{1.0}
\plotone{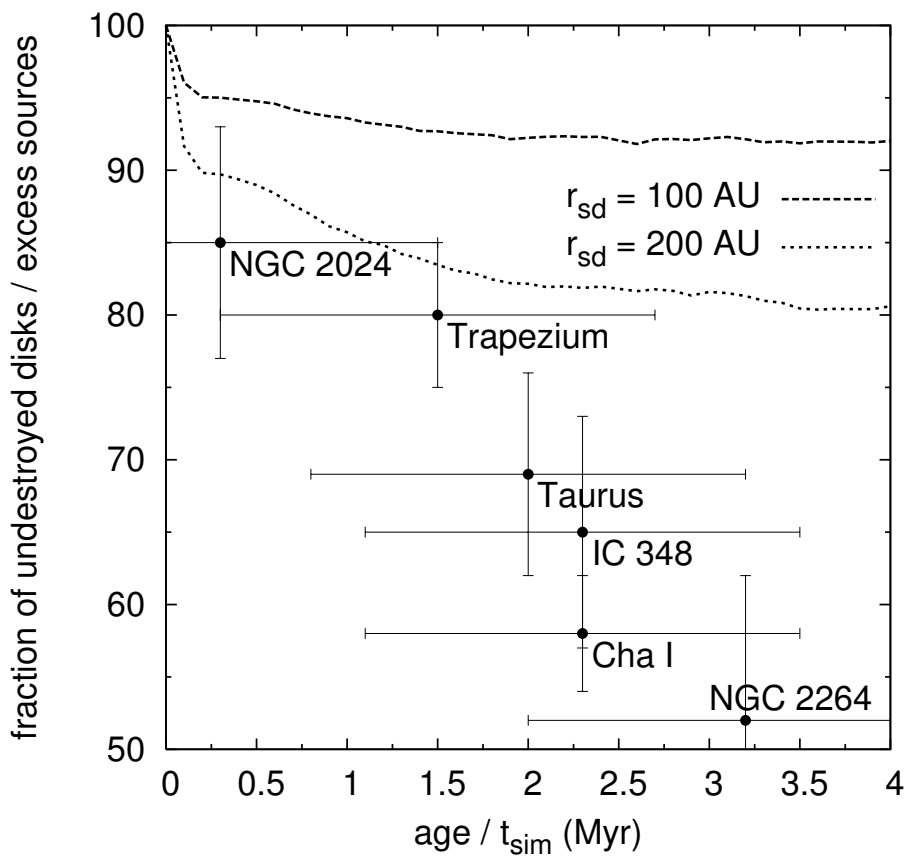}
\caption{Fraction of un-destroyed discs (i.e. discs with at least 10\% of
  their initial mass) in the Trapezium cluster for two different standard disc
  sizes, $r_{\rm{sd}}=100\,$AU (dashed line) and $r_{\rm{sd}}=200\,$AU (dotted
  line), model~A (cf. Fig.~\ref{losstime}). The simulation results are
  compared to the disc fraction of different star clusters, estimated from
  JHKL excess. The data were taken from \cite{haisch:apj01}.
\label{discfrac}}
\end{figure}




\clearpage
\begin{table}
\begin{center}
\begin{tabular}{rrrrrrrr}
\tableline    \tableline
\tableline
\\[-2ex]
$N$       & $R_{\rm{ONC}}$ & $\rho(r)$         & $\rho_{\rm{core}}$      & $\sigma_{\rm{3D}}$ & $\xi(m)$ & $Q_{\rm{vir}}$ & $t_{\rm{ONC}}$\\
          & [pc]      &                   & [$\mbox{pc}^{-3}$] & [km\,s$^{-1}$] &          &           & [Myr]\\
\\[-2ex]
\hline
\\[-2ex]
$\ge4000$ & 2.5       & $\propto{r^{-2}}$ & $4.7\times10^4$    & $4.3\pm0.5$   & KTG93    & ${1.5\,?}$ & $\approx2$\\
\\[-2ex]
\tableline
\end{tabular}
\caption{Main properties of the ONC.\label{table:PropONC}} 
\end{center}
\end{table}

\begin{table}
\begin{center}
\begin{tabular}{rrrrrrrrrrr}
\tableline    \tableline
\tableline
\\[-2ex]
Model      & $t_{\rm{out}}$  & $Q_{\rm{vir}}$  & $N$  & $N_{\rm{ONC}}$  & $R$   & $R_{\rm{hm}}$ & $\rho_{\rm{core}}$  & $\sigma_{\rm{3D}}$  & $t_{\rm{cross}}$ & $t_{\rm{relax}}$ \\ 
           & [Myr]              &                    &      &                    & [pc]  & [pc]              & [$10^4\,\mbox{pc}^{-3}$]  & [km\,s$^{-1}$]& [Myr]                & [Myr]\\
\\[-2ex]
\hline
\\[-2ex]
{\bf A} & 0                  & 0.50               & 4000 & 4000               & ~1.10 & 0.55              & 27.6                      & 4.11            & 0.4      & 24\\
           & 1                  & 0.63               & 3999 & 3962               & ~7.18 & 0.50              & ~8.0                      & 4.54            &\\
           & 2                  & 0.61               & 3998 & 3717               & 11.85 & 0.65              & ~3.8                      & 3.87            &\\
\\[-2ex]
\bf{A}$^*$ & 0                  & 0.50               & 4000 & 4000               & ~2.00 & 0.99              & 13.8                      & 3.05            & 0.6      & 36\\
           & 1                  & 0.60               & 4000 & 3902               & ~4.99 & 0.87              & ~3.6                      & 3.58            &\\
           & 2                  & 0.62               & 3999 & 3647               & 11.46 & 0.89              & ~1.5                      & 3.44            &\\
\\[-2ex]
{\bf B} & 0                  & 1.00               & 6000 & 6000               & ~0.70 & 0.34              & 68.5                      & 8.88            & 0.1      & 9\\
           & 1                  & 1.11               & 5999 & 4986               & ~9.33 & 1.07              & ~4.4                      & 5.79            &\\
           & 2                  & 1.36               & 5999 & 3762               & 20.22 & 1.40              & ~2.1                      & 5.52            &\\
\\[-2ex]
{\bf C} & 0                  & 0.10               & 5000 & 4167               & ~3.00 & 1.49              & ~7.4                      & 1.23            & 2.5      & 180\\
           & 1                  & 0.33               & 5000 & 4201               & ~3.24 & 1.45              & 10.6                      & 2.49            &\\
           & 2                  & 0.46               & 5000 & 4316               & ~5.00 & 1.34              & ~8.0                      & 3.06            &\\
\\[-2ex]
ONC        & -                  & 1.5 (?)            & -    & $\sim$ 4000        & -     & 0.5--0.8          & ~4.7                      & 4.3             & 0.5      & ~30 \\
\\[-2ex]
\tableline
\end{tabular}
\caption{Important quantities for the three most satisfying cluster models at different simulation times, 
               compared to the ONC model, where $t_{\rm{out}}$ is the time of data output, $Q_{\rm{vir}}$ the virial ratio, 
               $N$ the total number of particles, $N_{\rm{ONC}}$ the number of particles 
               inside 2.5~pc, $R$ the total radius of cluster,  $R_{\rm{hm}}$ the half-mass radius, $\rho_{\rm{core}}$ 
               the stellar density of the inner core with radius 0.053~pc, $\sigma_{\rm{3D}}$ the three-dimensional 
               velocity dispersion, $t_{\rm{cross}}$ the crossing time and $t_{\rm{relax}}$ the relaxation time.\label{table:sim2}}
\end{center}
\end{table}

\clearpage
\begin{table}
\begin{scriptsize}
\begin{center}
\begin{tabular}{r||*{14}{r}}
 & 500.0 & 90.0 & 50.0 & 20.0 & 9.0 & 5.0 & 4.0 & 3.0 & 2.0 & 1.5 & 1.0 & 0.5 & 0.3 & 0.1\\[0.5ex]
\tableline
\tableline
\\[-2ex]
0.1 & 0.930 & 0.899 & 0.879 & 0.890 & 0.862 & 0.799 & 0.737 & 0.713 & 0.694 & 0.680 & 0.631 & 0.599 & 0.423 & 0.167\\[0.5ex]
0.2 & 0.950 & 0.940 & 0.929 & 0.932 & 0.919 & 0.889 & 0.868 & 0.850 & 0.829 & 0.787 & 0.658 & 0.561 & 0.396 & 0.161\\[0.5ex]
0.3 & 0.979 & 0.983 & 0.982 & 0.965 & 0.950 & 0.930 & 0.913 & 0.876 & 0.831 & 0.786 & 0.700 & 0.528 & 0.349 & 0.155\\[0.5ex]
0.4 & 0.987 & 0.987 & 0.987 & 0.981 & 0.965 & 0.931 & 0.911 & 0.886 & 0.819 & 0.764 & 0.668 & 0.489 & 0.291 & 0.140\\[0.5ex]
0.5 & 0.991 & 0.989 & 0.987 & 0.979 & 0.941 & 0.898 & 0.880 & 0.846 & 0.799 & 0.740 & 0.665 & 0.475 & 0.255 & 0.138\\[0.5ex]
0.7 & 0.990 & 0.986 & 0.983 & 0.941 & 0.883 & 0.822 & 0.798 & 0.752 & 0.690 & 0.655 & 0.580 & 0.432 & 0.225 & 0.119\\[0.5ex]
1.0 & 0.989 & 0.969 & 0.938 & 0.873 & 0.781 & 0.694 & 0.660 & 0.615 & 0.550 & 0.500 & 0.433 & 0.284 & 0.175 & 0.090\\[0.5ex]
1.5 & 0.980 & 0.898 & 0.846 & 0.742 & 0.613 & 0.517 & 0.475 & 0.419 & 0.340 & 0.293 & 0.212 & 0.118 & 0.084 & 0.024\\[0.5ex]
2.0 & 0.955 & 0.827 & 0.759 & 0.619 & 0.469 & 0.352 & 0.308 & 0.248 & 0.175 & 0.125 & 0.089 & 0.037 & 0.019 & 0.000\\[0.5ex]
2.5 & 0.915 & 0.751 & 0.664 & 0.493 & 0.336 & 0.218 & 0.181 & 0.139 & 0.089 & 0.052 & 0.023 & 0.001 & 0.000 & 0.000\\[0.5ex]
3.0 & 0.877 & 0.671 & 0.570 & 0.385 & 0.224 & 0.114 & 0.089 & 0.054 & 0.026 & 0.013 & 0.000 & 0.000 & 0.000 & 0.000\\[0.5ex]
3.5 & 0.831 & 0.601 & 0.487 & 0.286 & 0.132 & 0.060 & 0.040 & 0.019 & 0.001 & 0.000 & 0.000 & 0.000 & 0.000 & 0.000\\[0.5ex]
4.0 & 0.793 & 0.523 & 0.397 & 0.209 & 0.085 & 0.018 & 0.006 & 0.000 & 0.000 & 0.000 & 0.000 & 0.000 & 0.000 & 0.000\\[0.5ex]
4.5 & 0.740 & 0.461 & 0.318 & 0.133 & 0.033 & 0.002 & 0.000 & 0.000 & 0.000 & 0.000 & 0.000 & 0.000 & 0.000 & 0.000\\[0.5ex]
5.0 & 0.705 & 0.386 & 0.255 & 0.097 & 0.013 & 0.000 & 0.000 & 0.000 & 0.000 & 0.000 & 0.000 & 0.000 & 0.000 & 0.0\\[0.5ex]
5.5 & 0.654 & 0.310 & 0.190 & 0.048 & 0.000 & 0.000 & 0.000 & 0.000 & 0.000 & 0.000 & 0.000 & 0.000 & 0.0 & 0.0\\[0.5ex]
6.0 & 0.618 & 0.252 & 0.156 & 0.029 & 0.000 & 0.000 & 0.000 & 0.000 & 0.000 & 0.000 & 0.0 & 0.0 & 0.0 & 0.0\\[0.5ex]
6.5 & 0.575 & 0.211 & 0.104 & 0.012 & 0.000 & 0.000 & 0.000 & 0.000 & 0.000 & 0.0 & 0.0 & 0.0 & 0.0 & 0.0\\[0.5ex]
7.0 & 0.537 & 0.156 & 0.087 & 0.001 & 0.000 & 0.000 & 0.000 & 0.000 & 0.000 & 0.0 & 0.0 & 0.0 & 0.0 & 0.0\\[0.5ex]
7.5 & 0.492 & 0.132 & 0.047 & 0.000 & 0.000 & 0.000 & 0.0 & 0.0 & 0.0 & 0.0 & 0.0 & 0.0 & 0.0 & 0.0\\[0.5ex]
8.0 & 0.459 & 0.117 & 0.034 & 0.000 & 0.000 & 0.000 & 0.0 & 0.0 & 0.0 & 0.0 & 0.0 & 0.0 & 0.0 & 0.0\\[0.5ex]
8.5 & 0.418 & 0.079 & 0.011 & 0.000 & 0.000 & 0.0 & 0.0 & 0.0 & 0.0 & 0.0 & 0.0 & 0.0 & 0.0 & 0.0\\[0.5ex]
9.0 & 0.373 & 0.071 & 0.005 & 0.000 & 0.000 & 0.0 & 0.0 & 0.0 & 0.0 & 0.0 & 0.0 & 0.0 & 0.0 & 0.0\\[0.5ex]
9.5 & 0.340 & 0.038 & 0.001 & 0.000 & 0.0 & 0.0 & 0.0 & 0.0 & 0.0 & 0.0 & 0.0 & 0.0 & 0.0 & 0.0\\[0.5ex]
10.0 & 0.291 & 0.026 & 0.000 & 0.000 & 0.0 & 0.0 & 0.0 & 0.0 & 0.0 & 0.0 & 0.0 & 0.0 & 0.0 & 0.0\\[0.5ex]
10.5 & 0.265 & 0.010 & 0.000 & 0.000 & 0.0 & 0.0 & 0.0 & 0.0 & 0.0 & 0.0 & 0.0 & 0.0 & 0.0 & 0.0\\[0.5ex]
11.0 & 0.229 & 0.005 & 0.000 & 0.000 & 0.0 & 0.0 & 0.0 & 0.0 & 0.0 & 0.0 & 0.0 & 0.0 & 0.0 & 0.0\\[0.5ex]
11.5 & 0.216 & 0.000 & 0.000 & 0.0 & 0.0 & 0.0 & 0.0 & 0.0 & 0.0 & 0.0 & 0.0 & 0.0 & 0.0 & 0.0\\[0.5ex]
12.0 & 0.176 & 0.000 & 0.000 & 0.0 & 0.0 & 0.0 & 0.0 & 0.0 & 0.0 & 0.0 & 0.0 & 0.0 & 0.0 & 0.0\\[0.5ex]
12.5 & 0.168 & 0.000 & 0.000 & 0.0 & 0.0 & 0.0 & 0.0 & 0.0 & 0.0 & 0.0 & 0.0 & 0.0 & 0.0 & 0.0\\[0.5ex]
13.0 & 0.134 & 0.000 & 0.000 & 0.0 & 0.0 & 0.0 & 0.0 & 0.0 & 0.0 & 0.0 & 0.0 & 0.0 & 0.0 & 0.0\\[0.5ex]
13.5 & 0.130 & 0.000 & 0.000 & 0.0 & 0.0 & 0.0 & 0.0 & 0.0 & 0.0 & 0.0 & 0.0 & 0.0 & 0.0 & 0.0\\[0.5ex]
14.0 & 0.098 & 0.000 & 0.000 & 0.0 & 0.0 & 0.0 & 0.0 & 0.0 & 0.0 & 0.0 & 0.0 & 0.0 & 0.0 & 0.0\\[0.5ex]
14.5 & 0.098 & 0.000 & 0.000 & 0.0 & 0.0 & 0.0 & 0.0 & 0.0 & 0.0 & 0.0 & 0.0 & 0.0 & 0.0 & 0.0\\[0.5ex]
15.0 & 0.069 & 0.000 & 0.000 & 0.0 & 0.0 & 0.0 & 0.0 & 0.0 & 0.0 & 0.0 & 0.0 & 0.0 & 0.0 & 0.0\\[0.5ex]
15.5 & 0.070 & 0.000 & 0.0 & 0.0 & 0.0 & 0.0 & 0.0 & 0.0 & 0.0 & 0.0 & 0.0 & 0.0 & 0.0 & 0.0\\[0.5ex]
16.0 & 0.071 & 0.000 & 0.0 & 0.0 & 0.0 & 0.0 & 0.0 & 0.0 & 0.0 & 0.0 & 0.0 & 0.0 & 0.0 & 0.0\\[0.5ex]
16.5 & 0.047 & 0.000 & 0.0 & 0.0 & 0.0 & 0.0 & 0.0 & 0.0 & 0.0 & 0.0 & 0.0 & 0.0 & 0.0 & 0.0\\[0.5ex]
17.0 & 0.042 & 0.000 & 0.0 & 0.0 & 0.0 & 0.0 & 0.0 & 0.0 & 0.0 & 0.0 & 0.0 & 0.0 & 0.0 & 0.0\\[0.5ex]
17.5 & 0.023 & 0.0 & 0.0 & 0.0 & 0.0 & 0.0 & 0.0 & 0.0 & 0.0 & 0.0 & 0.0 & 0.0 & 0.0 & 0.0\\[0.5ex]
18.0 & 0.021 & 0.0 & 0.0 & 0.0 & 0.0 & 0.0 & 0.0 & 0.0 & 0.0 & 0.0 & 0.0 & 0.0 & 0.0 & 0.0\\[0.5ex]
18.5 & 0.010 & 0.0 & 0.0 & 0.0 & 0.0 & 0.0 & 0.0 & 0.0 & 0.0 & 0.0 & 0.0 & 0.0 & 0.0 & 0.0\\[0.5ex]
19.0 & 0.008 & 0.0 & 0.0 & 0.0 & 0.0 & 0.0 & 0.0 & 0.0 & 0.0 & 0.0 & 0.0 & 0.0 & 0.0 & 0.0\\[0.5ex]
19.5 & 0.002 & 0.0 & 0.0 & 0.0 & 0.0 & 0.0 & 0.0 & 0.0 & 0.0 & 0.0 & 0.0 & 0.0 & 0.0 & 0.0\\[0.5ex]
20.0 & 0.001 & 0.0 & 0.0 & 0.0 & 0.0 & 0.0 & 0.0 & 0.0 & 0.0 & 0.0 & 0.0 & 0.0 & 0.0 & 0.0\\[0.5ex]
\end{tabular}
\caption{Table of relative disk mass losses $\Delta{M}_{\rm{d}}/M_{\rm{d}}$ for all simulated configurations of parabolic $(e=1)$ star-disc encounters. The first row contains the relative perturber masses $M_2^*/M_1^*$, the first column contains the relative periastra $r_{\rm{p}}/r_{\rm{d}}$. Results from simulations are denoted by four digits, the values ``0.0'' were edited manually.\label{table:massloss}}
\end{center}
\end{scriptsize}
\end{table}

\end{document}